\documentclass[preprint,12pt]{elsarticle}

\usepackage{graphicx}
\usepackage{amssymb}
\usepackage{amsmath}
\usepackage{color}
\usepackage{multirow}
\usepackage{subfigure}
\usepackage[margin=1.5in]{geometry}

\begin{document}
\title{A Kaluza-Klein Inspired Brans-Dicke Gravity with Dark Matter and Dark Energy Model}

\author[1,4]{Chakrit Pongkitivanichkul}
\ead{chakpo@kku.ac.th}
\author[1,4]{Daris Samart}
\ead{darisa@kku.ac.th}
\author[1,2]{Nakorn Thongyoi}
\ead{nakorn.thongyoi@gmail.com}
\author[3]{Nutthaphat Lunrasri}
\ead{n.lunrasri@gmail.com}
\address[1]{Khon Kaen Particle Physics and Cosmology Theory Group (KKPaCT),\\ Department of Physics, Faculty of Science, Khon Kaen University, 123 Mitraphap Rd., Khon Kaen, 40002, Thailand}
\address[2]{Department of Physics, University of Southampton, SO17 1BJ Southampton,
United Kingdom}
\address[3]{Department of Physics, Faculty of Science, Chulalongkorn University, Phyathai Rd., Bangkok 10330, Thailand}
\address[4]{National Astronomical Research Institute of Thailand, Chiang Mai 50180, Thailand}

\date{\today} 

\begin{abstract}
We propose the Kaluza-Klein inspired Brans-Dicke gravity model containing possible existence of dark matter and dark energy. The massive scalar field coupled with gravity in 5 dimensional spacetime can be reduced to 4 dimensional gravity along with the dilaton $\phi$, gauge fields $A_{\mu}$, and the tower of scalar fields $\eta_n$. Two additional gauge fields are introduced to form ``Cosmic Triad" vector field scenario. We then use the dynamical system approach to analyze the critical points and their corresponding physical parameters. We found that in the case where only the zero mode of the Kaluza-Klein scalar is decoupled, the system contains both dark matter and dark energy phase depending on the mass parameter with the presence of the gauge field.
\end{abstract}


\maketitle
\section{Indroduction} \label{sec:intro}
On the one hand, the mysterious accelerated expansion of the universe is one of the major problems in modern cosmology at the present. Recent observational data of supernovae type Ia indicates that the universe is expanding with acceleration at a large scale \cite{Riess:1998cb,Perlmutter:1998np}. This is the so-called “Dark Energy” (DE) problem. According to the observational data, we find that the universe is dominated by the DE which accounts for 70\% of the total energy density \cite{Bennett:2003bz,Halverson:2001yy}. There is no conventional physical mechanism capable of completely explaining the accelerated cosmic expansion. 
The simplest idea of introducing a cosmological constant seems to agree with the observational data of the accelerating universe. However the cosmological constant suffers from a serious problem in the theoretical sense, i.e., the values of the cosmological constant coming from field theory is extremely bigger than the value from the observational data 
\cite{Carroll:2000fy,Padmanabhan:2002ji}. 
Several alternative phenomenological suggestions and theoretical hypotheses have been proposed to resolve this issue instead. The most popular approach is to introduce a new form of matter or an exotic matter with negative pressure as a source of the accelerating universe \cite{Peebles:2002gy,Copeland:2006wr}. So far, however, there is no evidence or experiment to prove the existence of the exotic matter, yet.

On the other hand, the nature of Dark Matter (DM) is a part of the unsolved problems in physics 
\cite{Bertone:2016nfn}. DM was proposed as a hypothetical particle with no electromagnetic interaction but its gravitational interaction is responsible for holding galaxies and forming the large scale structure of the universe \cite{Zwicky:1933gu,Begeman:1991iy}. The observational results have been suggesting that DM composes about 27\% of the total energy density of the universe. This kind of matter interacts very weakly, if not at all, to all known fundamental forces in the standard model of particle physics (SM). 
Figuring out what DM is made of is one of current on-going particle physics research \cite{Bertone:2004pz,Jungman:1995df,Feng:2010gw}. 

These lead to reconsideration or a modification of GR at the large scale \cite{DeFelice:2010aj,Sotiriou:2008rp,Capozziello:2011et,Clifton:2011jh,Nojiri:2006ri,Capozziello:2002rd}. By retaining prominent features of GR, the modified gravity is aimed to solve several GR’s problems not only for DE problem, for instance, non-renormalizability, DM and so on. Notice that there are countless inequivalent ways to modify gravity leading to theories that can be designed to reconcile with current observations. Cosmological observations suggest that GR must be modified at very low and/or very high energies. Experimental searches for beyond-GR physics are a particularly active and well-motivated area of research. The simple extension of the modified gravity is to consider adding an additional degree of freedom in the theories. This extension of modified gravity is equivalent to the scalar–tensor theories which have been extensively studied in the literature see \cite{Fujii:2003pa,Faraoni:2004pi} and references therein. The main idea of the scalar-tensor theories is that the gravitational interaction is mediated by scalar and tensor fields. The most popular and well-behaved model of the scalar-tensor theories is the Brans-Dicke (BD) theory. Inspired by Mach’s principle, the reference frame of the BD theory comes from the distribution of matters in the universe \cite{Brans:1961sx}. In GR, geometry is determined by mass distribution. However, it is not unique up to boundary condition. This requires the gravitational constant of the BD theory depending on space-time. This means that the gravitational constant plays the role of a dynamical scalar field $\phi$. In addition, the BD theory is proposed to make the results of GR compatible with several experimental and observational data from small to large scales. For example, observations of the solar system and other gravitational systems constrained the parameter of the BD theory, $\omega$ as $\omega > 40,000$ \cite{Will:2018bme} where the theory approaches to GR when $\omega\to \infty$. 

The BD theory has been studied in order to solve various problems in gravity and cosmology. 
The dynamical system method has been utilized as an useful tool for investigating several aspects in cosmology \cite{Wainwright:1997,Coley:2003mj} and see \cite{Bahamonde:2017ize} for recent review. In particular, it is used to gain a qualitative understanding of the dynamics of the universe in asymptotic regions, i.e., early and late times of the universe. Having use of the dynamical system analysis, a good candidate of the cosmological models needs to consequently evolve following the series of standard cosmological epochs, i.e., inflation $\to$ radiation $\to$ dust (matter) $\to$ DE phase \cite{Copeland:2006wr}. A huge number of the studies in the BD cosmological models are extensively analyzed using dynamical system in various classes of the potentials with additional matter fields \cite{Kolitch:1994kr,Santos:1996jc,Copeland:1997et,Abdalla:2007qi,deSouza:2005ig,Hrycyna:2013hla,Hrycyna:2013yia,Hrycyna:2014cka,Garcia-Salcedo:2015naa,Papagiannopoulos:2016dqw,Felegary:2016znh,Roy:2017mnz,Ghaffarnejad_2017,Lu:2019twd,Shabani_2019,Zucca:2019ohv,Giacomini:2020grc}. Interestingly, the BD gravity with the quadratic potential gives the de-Sitter (dS) attractor critical point in the phase space of this model and provides an explanation of the accelerated expansion of the Universe without introducing any form of the exotic matter fields \cite{Hrycyna:2013yia}. Nevertheless, choosing the types of the potential term in the BD gravity is somewhat arbitrary and a particular form of the potential does not occur naturally from the BD model. We will show in the latter that using only extra fields the same cosmological consequences can be achieved.

Moreover, a common prediction from the string theory, one of the candidate of quantum gravity theories with the extra-dimension, is an existence of the scalar field (spin-0) a counterpart of the graviton (spin-2) called the dilaton \cite{Zee:1978wi}. Therefore, at low-energies, string theory suggests that the compatible theory of gravity is the scalar-tensor gravity rather than GR. The dilaton plays a central role in several observed phenomena in high energy physics such as the inflaton field \cite{Gasperini:1993hu,Damour:1995pd,Chamblin:1999ya}, DM \cite{Damour:1990tw}, Dark Radiation (DR) \cite{Svrcek:2006yi,Acharya:2015zfk}, and cosmological constant  \cite{Wetterich:1987fm,Shaposhnikov:2008xi}.
Originally, the Kaluza Klein (KK) theory was first introduced to combine gravity (GR) and electromagnetic theory in the single framework by introducing extra dimension see review and monograph \cite{Appelquist:1987nr,Overduin:1998pn,Wesson:2006ta,Wesson:2019rcb}. The idea of compactification of higher dimension has been a corner stone of the string theory since \cite{Bailin:1987jd,Duff:1986hr}. 

Inspired by the KK theory, we will show that the BD theory with a particular value of the $\omega$ parameter is obtainable by the KK dimensional reduction process and the residual fields are perfect candidates for solving DM \& DE problems. This work is organized as follow, the simple toy model of the KK inspired BD gravity is set up in section 2. Next section, we derive all relevant equations of motion. In section 4, we restrict our model in a lower mass case of the KK inspired BD. To demonstrate a possibility of DM \& DE existence in our model, the autonomous equations in the light of the dynamical system framework is derived and the stability of the fixed points are analyzed in section 5. The results and discussion are in section 6. We close this work by conclusion section with some implication of this work in the last section.    
\section{The Model} \label{sec:model}
The KK-inspired BD model will be derived in this section. We start with the 5 dimensional action for KK gravity and a free massive scalar field, $\widetilde{\eta}$ propagating in the 5 dimensional spacetime, it reads,
\begin{equation}
S = \int d^4 x dy \sqrt{-\widetilde{g}}\, \sum_{a = 1}^{2}\left( -\widetilde{\mathcal{R}} + \hat \partial_A \widetilde{\eta}^* \partial^A \widetilde{\eta} - M_{(5)}^2 \widetilde{\eta}^* \widetilde{\eta} -\frac{1}{4}\phi^2 \widetilde{F}^a_{AB}\widetilde{F}^{aAB} + V(\widetilde{A}_M^{a})\right)
\end{equation}
where we used Planck mass unit, $\sqrt{16\pi G} = 1$, and the metric convention is mostly minus. The mass parameter of the scalar field in 5 dimension spacetime is denoted as $M_{(5)}$. Two additional gauge fields are denoted by $\widetilde{F}_{MN}^a = \partial_M \widetilde{A}^a_N - \partial_N \widetilde{A}^a_M$ where $a = 1,2$. In this work, we have used the capital Roman alphabets representing the 5 dimensional spacetime indices as $A,B,C, \cdots = 0,1,2,3,5$ while the Greek alphabets stand for physical dimensional spacetime indices via $\mu,\nu,\sigma,\cdots = 0,1,2,3$. The $\mathcal{\widetilde{R}}= \mathcal{\widetilde{R}}_{AB}\,\widetilde{g}^{AB}$ is the 5 dimensional Ricci scalar and the 5 dimensional spacetime metric, $\widetilde{g}_{AB}$ is given by \cite{Overduin:1998pn},
\begin{equation}
\widetilde{g}_{AB} = \begin{pmatrix} g_{\mu\nu} + \phi^2A_{\mu}A_{\nu} & \phi^2A_{\mu}\\
\phi^2A_{\nu} & \phi^2
\end{pmatrix}\,,
\end{equation}
where $A^\mu$ is the electromagnetic gauge field and $\phi$ is the dilaton field. Assuming that the extra dimension is compactified in a circle of radius $R_k$, there is a periodic shift symmetry in the 5th direction as
\begin{equation}
y \rightarrow y + 2\pi R_k.
\end{equation}
After performing a dimensional reduction, the gravitational part of the action becomes
\begin{equation}
S_{BD} = -\int d^4 x \sqrt{-g}\phi\left( \mathcal{R} + \frac{1}{4}\phi^2F_{\mu\nu}F^{\mu\nu} + \frac{2}{3}\frac{\partial^{\mu}\phi \partial_{\mu}\phi}{\phi^2}\right),
\label{BD-action}
\end{equation}
The above action is equivalent to the BD gravity with $\omega=-4/3$. We note that the KK gravity with the dimensional reduction in 5 dimension gives a non-minimal coupling to the 4 dimensional Ricci scalar which is equivalent to the BD gravity with non-minimal coupling to the EM field. The action for the free scalar field can also be reduced to 4 dimensional action as follow. Starting with the Fourier expansion of the complex scalar field, 
\begin{eqnarray}
\widetilde{\eta}(x^{\mu},y) = \sum_{n=0}^{\infty} e^{i{y_n}/{R_k}}\hat{\eta}_n(x^{\mu}),
\end{eqnarray}
then integration over the extra dimension gives
\begin{equation}        
\int_0 ^{2\pi R_k}dy \partial_M \widetilde{\eta}^* \partial^M \widetilde{\eta} = 2\pi R_k \sum_{n=0}^{\infty} \left[ \partial_{\mu}\hat{\eta}_n \partial^{\mu}\hat{\eta}_n^* + \left(\frac{1}{\phi^2} + \kappa^2 A_{\nu}A^{\nu}\right) \frac{n^2}{R_k^2}\hat{\eta}_n\hat{\eta}_n^* \right].
\end{equation}
We redefine the new field with    
\begin{equation}
\eta_n = \sqrt{2\pi R_k} \hat{\eta}_n.
\label{fourier}
\end{equation}
Using the assumption that the 4 dimensional fields are real, i.e., $\eta_n^* = \eta_n$, the action becomes canonical scalar field which is the starting point for us to solve the equation of motion:
\begin{equation}
S_{KK} = \int d^4 x \sqrt{-g}\phi \sum_{n=0}^{\infty}\left[ \partial_{\mu}\eta_n \partial^{\mu}\eta_n - M_{(5)}^2 \eta^2 + \left(\frac{1}{\phi^2} + A_{\nu}A^{\nu}\right)\frac{n^2}{R_k^2}\eta_n^2 \right].
\label{4d-scalar-KK}
\end{equation}

The dimensional reduction of the gauge field part can be done similarly. Fourier transformation of gauge fields is written as
\begin{equation}
    \widetilde{A}^a_M = \sum_{n=0}^{\infty} e^{iy_n/R_k}\widetilde{A}_{M,n}^{a}(x^{\mu}).
\end{equation}
Requiring the gauge fields to be real, and redefine the field with the factor $\sqrt{2\pi R_k}$. The integration over the extra dimension becomes
\begin{align}
    S_G = \int d^4x \sqrt{-g} \phi \sum_{a=1}^{2}\sum_{n=0}^{\infty} \left( -\frac{1}{4}\phi^2\widetilde{F}^a_{n\mu\nu}\widetilde{F}^{a\mu\nu}_n + \left( V'' - \frac{\phi^2 n^2}{4 R_k^2} \right) \widetilde{A}^a_{\mu,n} \widetilde{A}^{\mu,a}_n \right. & \nonumber \\
    \left. -\frac{1}{4}\phi^2 \partial_{\mu} \widetilde{A}_{5,n}^a \partial^{\mu}\widetilde{A}^{5,a}_n + V''_{(5)}\widetilde{A}^a_{5,n} \widetilde{A}^{5,a}_n \right), &
\end{align}
where $V'' = \frac{\delta^2 V}{\delta A_{\mu}^2}$ and $V''_{(5)} = \frac{\delta^2 V}{\delta A_{5}^2}$. 

Before we move forward to calculate the equations of motion in the next section. It is worth to remark some discussions about the KK inspired BD gravity in the present work. One sees that the dilation is generated by the dimensional reduction from the compactification of the extra (fifth) dimension in the KK gravity and non-minimally couples to the gravity in 4 dimension. Moreover, having use the Fourier expansion of the complex scalar field in Eq. (\ref{fourier}), this leads to a certain form of the potential from the KK inspired BD model that the dilaton is coupled with the scalar fields, $\eta_{n}$, from the 5-dimension as shown in Eq. (\ref{4d-scalar-KK}). This is a salient feature of the KK inspired BD gravity with a specific form of the potential whereas the potential form of the traditional BD graviy model is arbitrary.
\section{Equation of Motions} \label{sec:eom}
In this section, we are going to work out the equations of motion in our model that are useful in the following when the equations of the dynamical system are derived.  We would like to study the the FRW universe with the spatially flat that satisfies the metric 
\begin{eqnarray}
    ds^{2}=dt^{2}-a^{2}(t)(dx^{2}+dy^{2}+dz^{2}),
\end{eqnarray} so $g_{tt}=1$ and $g_{ij}=-\delta_{ij}a^{2}(t)$ for $i,j=1,2,3$. where $a(t)$ is the scale factor in the co-moving frame.\\
    \indent The non-vanishing components of the Christoffel symbol in Cartesian coordinates are:
    \begin{equation}
    \Gamma_{ij}^t = \dot{a}a \delta_{ij},\;\;\;\Gamma^i_{tj} = \frac{\dot{a}}{a} \delta^i_j \,.
    \end{equation} 
    After deriving through the Euler Lagrange equation, we obtain the equation of motion for $\phi$ as:
    \begin{eqnarray}
        \mathcal{R} &=& \frac{4}{3\sqrt{-g}\phi} \partial^{\mu}(\sqrt{-g}\partial_{\mu}\phi) - \frac{3}{4}\phi^2 F_{\mu\nu} F^{\mu\nu} - \frac{2}{3\phi^2}\partial^{\mu}\phi\partial_{\mu}\phi \nonumber\\
        & & + \sum_{n=0}^{\infty}\left[ \partial_{\mu}\eta_n \partial^{\mu}\eta_n - M_{(5)}^2 \eta^2 + \left(\frac{1}{\phi^2} + A_{\nu}A^{\nu}\right)\frac{n^2}{R_k^2}\eta_n^2 \right] \nonumber\\
        & & +  \sum_{a=1}^{2}\sum_{n=0}^{\infty} \left( -\frac{3}{4}\phi^2\widetilde{F}^a_{n\mu\nu}\widetilde{F}^{a\mu\nu}_n + \left( V'' - \frac{3}{4}\frac{\phi^2 n^2}{ R_k^2} \right) \widetilde{A}^a_{\mu,n} \widetilde{A}^{\mu,a}_n \right) 
        ,
    \end{eqnarray}
    where we assume that gauge fields in 5th component are vanished, i.e., $\widetilde{A}^a_{5,n} = 0$. The equation of motion for the gauge field {from higher dimensional metric}, $A_{\mu}$ is given by:
    \begin{equation}
        \phi^3\nabla^{\mu}F_{\mu\nu} = - A_{\nu}\left(\sum_{n=0}^{\infty} \frac{n^2\eta^2_n}{R_k^2}\right),
    \end{equation}
    and the additional gauge fields, $\widetilde{A}_{\mu,n}^a$:
    \begin{equation}
        \phi^3\nabla^{\mu}\widetilde{F}_{\mu\nu,n}^a = - \widetilde{A}_{\nu,n}^a \sum_{n=0}^{\infty} \left( V'' -  \frac{n^2\phi^2}{4 R_k^2}\right),
    \end{equation}
    and the equation of motion for scalar field, $\eta_n$'s:
    \begin{equation}
    \frac{1}{\sqrt{-g}} \partial^{\mu}(\sqrt{-g}\partial_{\mu}\eta_n) = \phi \left(\frac{1}{\phi^2} + A_{\mu}A^{\mu}\right)\frac{n^2}{R_k^2} \eta_n - \phi M_{(5)}^2\eta_n.
    \end{equation}
    The energy-momentum tensor is defined as 
    \begin{eqnarray}
        T^{\mu\nu}&=&
        -\frac{2}{\sqrt{-g}}\frac{\delta(\sqrt{-g}\mathcal{L}_{\text{matter}})}{\delta g^{\mu\nu}},
        \label{Energy-momentum tensor}
    \end{eqnarray}
    For simplicity, we identify the matter field Lagrangian as
    \begin{eqnarray}
        \mathcal{L}_{\text{matter}}&=&-\frac{1}{4}\phi^{3}F^{2}-\frac{2}{3}\frac{\partial_{\mu}\phi\partial^{\mu}\phi}{\phi}\nonumber \\
        & & + \phi \left[ \partial_{\mu}\eta_n \partial^{\mu}\eta_n - M_{(5)}^2 \eta^2 + \left(\frac{1}{\phi^2} + A_{\nu}A^{\nu}\right)\frac{n^2}{R_k^2}\eta_n^2 \right].
        \label{Lagrangian for matter}
    \end{eqnarray}
    Let's put this into Eq.\eqref{Energy-momentum tensor}. We obtain the energy momentum tensor in the following form,
    \begin{eqnarray}
        T_{\mu\nu}&=&\phi^{3}F_{\mu\alpha}F_{\nu}\!^{\alpha}-\frac{g_{\mu\nu}}{4}\phi^3 F^2 + \frac{2}{3}\frac{\partial_{\mu}\phi\partial_{\nu}\phi}{\phi} - \phi \sum_{n=0}^{\infty} \partial_{\mu}\eta_{n}\partial_{\nu}\eta_{n} \\
        & & - 2 \phi A_{\mu}A_{\nu}\sum_{n=0}^{\infty} \frac{n^2\eta_n^2}{R_k^2} + g_{\mu\nu} \phi \sum_{n=0}^{\infty}\left[\left(\frac{1}{\phi^2} +  A_{\nu}A^{\nu}\right)\frac{n^2}{R_k^2}\eta_n^2 - M_{(5)}^2 \eta_n^2\right] \nonumber \\
        & & + \phi^3 \sum_{n=0}^{\infty} \widetilde{F}^a_{\mu\alpha,n} \widetilde{F}^{a,n}_{\nu} - \phi \sum_{n=0}^{\infty} \left( V'' - \frac{n^2 \phi^2}{4 R_k^2} \right) (2 \widetilde{A}^a_{\mu,n} \widetilde{A}^{a}_{\nu,n} - g_{\mu\nu} \widetilde{A}^a_{\rho,n} \widetilde{A}^{\rho,a}_{n}). \nonumber
    \end{eqnarray}
   In order to simplify the gauge fields sector, we will assume that only zero mode ($n = 0$) of each species ($a = 1,2$) has non-vanishing spatial components. The Cosmic Triad \cite{ArmendarizPicon:2004pm,Golovnev:2008cf,Maleknejad:2012fw} solution requires that
   \begin{eqnarray}
       A_{\mu} &=& (0,A,0,0)\\
       \widetilde{A}^1_{\mu,0} \equiv \widetilde{A}^1_{\mu}  &=& (0,0,A,0)\\
       \widetilde{A}^2_{\mu,0} \equiv \widetilde{A}^2_{\mu} &=& (0,0,0,A),
   \end{eqnarray}
   such that the off-diagonal components of the energy momentum tensor vanish. In order to make the diagonal spatial components 
   equal, we assume that
   \begin{equation}
       V'' = \sum_{n=0}^{\infty} \frac{n^2\eta_n^2}{R_k^2} \equiv M_A^2.
   \end{equation}
    
    For the calculation of the field strength tensor, $F^{\mu\nu}$, we assume homogeneous gauge and scalar fields conditions where
    \begin{equation}
    \partial_i A = 0,\quad {\rm and}\quad  \partial_i \phi = 0
    \end{equation}
    Therefore, the components can be calculated as follow:
    \begin{align}
    F^{ik} &= \widetilde{F}^{tk,1} = \widetilde{F}^{tk,2} = 0\\
    F^{ti} &= (\dot{A} + 2HA)\delta^{i1}\\
    \widetilde{F}^{ti,1} &= (\dot{A} + 2HA)\delta^{i2}\\
    \widetilde{F}^{ti,2} &= (\dot{A} + 2HA)\delta^{i3}
    \end{align}
    where above we use $H = \frac{\dot{a}}{a}$.

    Having use all relevant calculations, the Einstein field equation of the KK inspired BD gravity takes the form
    \begin{equation}
        \phi(\mathcal{R}_{\mu\nu}-\frac{1}{2}g_{\mu\nu}\mathcal{R}) + g_{\mu\nu}\nabla_{\sigma}\nabla^{\sigma}\phi - \nabla_{\mu}\nabla_{\nu}\phi = T_{\mu\nu}.
        \label{EFE}
    \end{equation}
    We can arrive at the Friedmann equation by looking for the $tt$ component in Einstein equation. The time component of the Einstein tensor is given by
    \begin{equation}
        G_{tt}=\frac{\dot{a}^{2}}{a^{2}},
    \end{equation}
    and the Friedmann equation becomes
    \begin{eqnarray}
        \left(\frac{\dot{a}}{a}\right)^{2}&=& -\frac{3\phi^2a^2}{2}\left(\dot{A} + 2 H A\right)^2 +\frac{2\dot{\phi}^2}{3\phi^2} - \sum_{n=0}^{\infty}\dot{\eta}_n^2 \nonumber \\
        & & + \sum_{n=0}^{\infty} \left[ \left(\frac{1}{\phi^2} -3 a^2 A^2\right)\frac{n^2}{R_k^2} - M_{(5)}^2 \right]\eta_n^2,
        \label{1st-Friedmann}
    \end{eqnarray}
    on the above calculation we use the fact that
    \begin{equation}
        \widetilde{F}^{t\sigma,1} \widetilde{F}^{t,1}\!_{\sigma} = \widetilde{F}^{t\sigma,2}\widetilde{F}^{t,2}\!_{\sigma} = F^{t\sigma}F^{t}\!_{\sigma} = g_{ij}F^{ti}F^{tj} = -a^2\left(\dot{A} + 2 H A\right)^2,
    \end{equation}
    and 
    \begin{equation}
        \widetilde{F}^{1}_{\mu\nu} \widetilde{F}^{\mu\nu,1} = \widetilde{F}^{2}_{\mu\nu} \widetilde{F}^{\mu\nu,2} =  F_{\mu\nu}F^{\mu\nu} = F_{ti}F^{ti} + F_{it}F^{it} = -2a^2\left(\dot{A} + 2 H A\right)^2.
    \end{equation}
    In order to calculate the Raychaudhuri's equation, we use the spatial components of Einstein field equation. The energy momentum tensor in the $ij$ components is
    \begin{eqnarray}
        T_{ij}&=&\phi^{3}a^4\left(\dot{A} + 2 H A\right)^2 \delta_{ij} -\frac{3 a^4 \delta_{ij}}{2}\phi^3 \left(\dot{A} + 2 H A\right)^2 \\
        &-& 2 \phi A^2 \delta_{ij} \sum_{n=0}^{\infty} \frac{n^2\eta_n^2}{R_k^2} - a^2 \delta_{ij}\phi \sum_{n=0}^{\infty}\left[\left(\frac{1}{\phi^2} - 3 a^2 A^2\right)\frac{n^2}{R_k^2}\eta_n^2 - M_{(5)}^2 \eta_n^2\right].\nonumber
    \end{eqnarray}
    We note that the energy-momentum tensor is proportional to $\delta_{ij}$ satisfying the assumption of the FRW metric, i.e., homogeneous and isotropy.
    From Einstein equation in Eq.(\ref{EFE}), the left-handed side is
    \begin{equation}
        \phi \mathcal{G}_{ij} + \frac{g_{ij}}{\sqrt{-g}}\partial_0 \left( \sqrt{-g} \partial^0 \phi \right) = -\phi \delta_{ij} \left(\dot{a}^2 + 2 a \ddot{a}\right) - \delta_{ij} \left(3a\dot{a}\phi + a^2 \ddot{\phi}\right)
    \end{equation}
    Using the relation $\dot{H} = \frac{\ddot{a}}{a} - \frac{\dot{a}^2}{a^2}$ and taking trace, we find
    \begin{eqnarray}
        - 3\ddot{\phi} &=& 3\phi H^2 + 6\phi \left(\dot{H} + H^2\right) + 9 H \dot{\phi} \nonumber \\
        &-& \frac{3 a^2\phi^3}{2} \left(\dot{A} + 2 H A\right)^2 - 6 \phi a^2 A^2 \sum_{n=0}^{\infty} \frac{n^2\eta_n^2}{R_k^2} \nonumber\\
        &-& 3\phi \sum_{n=0}^{\infty}\left[\left(\frac{1}{\phi^2} -3 a^2 A^2\right)\frac{n^2}{R_k^2}\eta_n^2 - M_{(5)}^2 \eta_n^2\right].
        \label{phi-ddot}
    \end{eqnarray}
In this section, we have derived all related equations of motion in the KK inspired BD gravity and they will be useful to derive the autonomous system in the dynamical system analysis below. Due to the complexity of the system of equations, we will focus our study only the vacuum case. We consider this work as a toy model and leave the inclusion of the barotropic fluid for the future work.
\section{Lower Mode Cases} \label{sec:low_n}
In KK theory, it is known that the eigenvalue of momentum operator in 5th direction of a higher mode $\eta_n$ is given by $|n|/R_k$. Therefore, a higher mode which has a momentum larger than our physical scale, i.e. reduced Planck scale, will be neglected. The remaining modes satisfying the condition:
\begin{equation}
n < R_k
\end{equation}
will play a role in equation of motions. In this project we will study the non-obvious simplest case, $1 < R_k < 2$. In this case, only zero mode and the first mode involve in the Lagrangian. The set of equations of motion becomes: for the dilaton field, $\phi$,
\begin{eqnarray}
\ddot{\phi} + 3 H \dot{\phi} &=& -\frac{9}{2}\phi\left(\dot{H} + 2H^2\right) - \frac{27}{8}\phi^3a^2 \mathcal{A}_c^2 + \frac{\dot{\phi}^2}{2\phi} - \frac{3}{4}\phi\dot{\eta}_0^2 - \frac{3}{4}\phi\dot{\eta}_1^2 \nonumber \\
& & + \frac{3}{4} \phi \left( \frac{1}{\phi^2} + 3 a^2 A^2 \right)\frac{\eta_1^2}{R_k^2} + \frac{3}{4}\phi M_{(5)}^2\eta_0^2 + \frac{3}{4}\phi M_{(5)}^2\eta_1^2 \label{eq:phiEOM2},
\end{eqnarray}
for the zero mode, $\eta_0$ and the first excited mode, $\eta_1$ scalar fields,
\begin{eqnarray}
\ddot{\eta}_0 + \frac{\dot{\phi}}{\phi}\dot{\eta}_0 + 3 H \dot{\eta}_0 + M_{(5)}^2\eta_0 &=& 0 ,\label{eq:eta0EOM2}\\
\ddot{\eta}_1 + \frac{\dot{\phi}}{\phi}\dot{\eta}_1 + 3 H \dot{\eta}_1 + M_{(5)}^2\eta_1 &=& \left(\frac{1}{\phi^2} - 3 a^2 A^2 \right)\frac{\eta_1}{R_k^2}.\label{eq:eta1EOM2}
\end{eqnarray}
The equations of motion for gauge fields are
\begin{equation} \dot{\mathcal{A}}_c + 3 H \mathcal{A}_c + \frac{ A \eta_1^2}{R_k^2} = 0, \label{eq:gaugeEOM2}
\end{equation}
where $\mathcal{A}_c \equiv \dot{A} + 2 H A$. Having use all relevant ingredients and substituting in Eq.(\ref{1st-Friedmann}), the final form of the Friedman equation is written as
\begin{eqnarray}
\left(\frac{\dot{a}}{a}\right)^{2} &=& -\frac{3 \phi^2a^2}{2}\mathcal{A}_c^2 +\frac{2\dot{\phi}^2}{3\phi^2} - \dot{\eta}_0^2 - \dot{\eta}_1^2 - M_{(5)}^2 \eta_0^2 - M_{(5)}^2 \eta_1^2 \nonumber \\
& & +  \left(\frac{1}{\phi^2} - 3 a^2 A^2\right) \frac{\eta_1^2}{R_k^2}. \label{eq:Friedman2}
\end{eqnarray}
Finally, the Raychaudhuri equation becomes

\begin{eqnarray}
- 3\ddot{\phi} &=& 3\phi H^2 + 6\phi \left(\dot{H} + H^2\right) + 9 H \dot{\phi} -\frac{3 a^2\phi^3}{2} \mathcal{A}_c^2 + 3 \phi a^2 A^2 \frac{\eta_1^2}{R_k^2} \nonumber \\
& & - \frac{3}{\phi}\frac{\eta_1^2}{R_k^2} + 3\phi M_{(5)}^2 \eta_0^2 + 3\phi M_{(5)}^2 \eta_1^2 .\label{eq:Ray2}
\end{eqnarray}

Solving Eqs.(\ref{eq:phiEOM2}) and (\ref{eq:Ray2}) together gives
\begin{eqnarray}
    \dot{H} &=& \frac{13  a^2 A^2 \eta_1^2}{10 R_k^2}-\frac{31}{20} a^2 \phi^2 \mathcal{A}_c^2-\frac{3}{10}  \dot{\eta}_1^2+\frac{7}{10} M_{(5)}^2  \eta_1^2-\frac{  \eta_1^2}{10 R_k^2 \phi^2}\nonumber \\
    & & -\frac{12 H^2}{5}-\frac{3}{10}  \dot{\eta}_0^2+\frac{7}{10} M_{(5)}^2  \eta_0^2+\frac{\dot{\phi}^2}{5 \phi^2},
    \label{Hdot}
\end{eqnarray}
and
\begin{eqnarray}
    \ddot{\phi} &=& -\frac{18  a^2 A^2 \eta_1^2 \phi}{5 R_k^2}+\frac{18}{5} a^2 \phi^3 \mathcal{A}_c^2+\frac{3}{5}   \phi \dot{\eta}_1^2-\frac{12}{5} M_{(5)}^2  \eta_1^2 \phi + \frac{6   \eta_1^2}{5 R_k^2 \phi}\nonumber \\
    & & -3 H \dot{\phi} + \frac{9}{5} H^2 \phi + \frac{3}{5}   \phi \dot{\eta}_0^2-\frac{12}{5} M_{(5)}^2  \eta_0^2 \phi - \frac{2 \dot{\phi}^2}{5 \phi}.
    \label{phiddot}
\end{eqnarray}
So far, we have derived all equations of motion which will be used in the following sections in order to perform a dynamical system analysis in KK inspired BD model. 
\section{Dynamical System} \label{sec:dynamic}
In this section, we will demonstrate an existence of DM and DE in KK inspired BD model by using the dynamical system method. The dynamical system is very suitable for qualitatively studying the dynamics of the universe. This might provide some hints of the evolution of the universe with matter fields. Next we develop the dynamical system by defining the set of parameters as follow:
\begin{eqnarray}
&X_1 = \frac{\sqrt{3}a}{H}\mathcal{A}_c,\quad X_2 = \sqrt{\frac{2}{3}}\frac{\dot{\phi}}{H\phi},\quad X_3 = \frac{\dot{\eta}_0}{H},\quad X_4 = \frac{\dot{\eta}_1}{H},\nonumber \\
&X_5 = \frac{1}{\phi},\quad X_6 = \sqrt{3} a A,\quad X_7 = \frac{\eta_1}{H R_k},\quad X_8 = \frac{\eta_0}{H R_k}.
\end{eqnarray}
Then the Friedmann equation can be used as a constrained equation
\begin{equation}
1 = -\frac{X_1^2}{2X_5^2} + X_2^2 - X_3^2 -  X_4^2 + \left(X_5^2 - X_6^2 \right)X_7^2 - M_{(5)}^2 R_k^2 \left(X_7^2 + X_8^2\right). \label{eq:const}
\end{equation}
Moreover, we define new parameters as
\begin{eqnarray}
\lambda \equiv M_{(5)}R_k \quad {\rm and} \quad \mu \equiv \frac{1}{H R_k}.
\label{lambda-mu}
\end{eqnarray}
Then the conformal derivative of new parameters can be derived in terms of newly defined parameters as follow:
\allowdisplaybreaks
\begin{eqnarray}
    \frac{1}{H}\frac{dX_1}{dt} &=& -\frac{3}{10} X_1 X_2^2+\frac{3}{10} X_1 X_3^2+\frac{3}{10}   X_1 X_4^2+\frac{1}{10}   X_1 X_5^2 X_7^2 + \frac{2 X_1}{5} -   X_6 X_7^2 \nonumber \\
    & & -\frac{13}{30}   X_1 X_6^2 X_7^2-\frac{7}{10}   \lambda^2 X_1 X_7^2-\frac{7}{10}   \lambda^2 X_1 X_8^2  + \frac{31 X_1^3}{60 X_5^2}, \label{eq:x1}
\\
    \frac{1}{H}\frac{dX_2}{dt} &=& \frac{31 X_1^2 X_2}{60 X_5^2}+\frac{2 \sqrt{6}   X_1^2}{5 X_5^2}+\frac{3}{10}    X_2 X_3^2+\frac{3}{10}    X_2 X_4^2+\frac{1}{10}    X_2 X_5^2 X_7^2\nonumber \\
    & & -\frac{13}{30}    X_2   X_6^2 X_7^2-\frac{7}{10} \lambda ^2    X_2 X_7^2-\frac{7}{10} \lambda ^2    X_2 X_8^2-\frac{1}{10} 3 X_2^3-\frac{7}{5} \sqrt{\frac{3}{2}} X_2^2 \nonumber \\
    & & -\frac{3 X_2}{5}+\frac{1}{5} \sqrt{6}    X_3^2+\frac{1}{5} \sqrt{6}    X_4^2+\frac{2}{5} \sqrt{6}    X_5^2 X_7^2-\frac{2}{5} \sqrt{6}      X_6^2 X_7^2\nonumber \\
    & &-\frac{4}{5} \sqrt{6} \lambda ^2    X_7^2-\frac{4}{5} \sqrt{6} \lambda ^2    X_8^2+\frac{3 \sqrt{6}}{5}, \label{eq:x2}
\\
    \frac{1}{H}\frac{dX_3}{dt} &=& \frac{31   X_1^2 X_3}{60 X_5^2}-\frac{3}{10} X_2^2 X_3-\sqrt{\frac{3}{2}} X_2 X_3+\frac{3}{10}    X_3 X_4^2+\frac{1}{10}    X_3 X_5^2 X_7^2\nonumber \\
    & & -\frac{13}{30}    X_3   X_6^2 X_7^2-\frac{7}{10} \lambda ^2    X_3 X_7^2-\frac{7}{10} \lambda ^2    X_3 X_8^2\nonumber \\
    & & +\frac{3}{10}    X_3^3-\frac{3 X_3}{5}-\lambda ^2 \mu X_8, \label{eq:x3}
\\
    \frac{1}{H}\frac{dX_4}{dt} &=& \frac{31   X_1^2 X_4}{60 X_5^2}-\frac{3}{10} X_2^2 X_4-\sqrt{\frac{3}{2}} X_2 X_4+\frac{3}{10}    X_3^2 X_4+\frac{1}{10}    X_4 X_5^2 X_7^2 \nonumber \\
    & & -\frac{13}{30}    X_4   X_6^2 X_7^2-\frac{7}{10} \lambda ^2    X_4 X_7^2-\frac{7}{10} \lambda ^2    X_4 X_8^2+\frac{3}{10}    X_4^3-\frac{3 X_4}{5}\nonumber \\
    & & +\mu X_5^2 X_7-\mu   X_6^2 X_7-\lambda ^2 \mu X_7, \label{eq:x4}
\\
    \frac{1}{H}\frac{dX_5}{dt} &=& -\sqrt{\frac{3}{2}} X_2 X_5, \label{eq:x5}
\\
    \frac{1}{H}\frac{d  X_6}{dt} &=&   X_1  -   X_6, \label{eq:x6}
\\
    \frac{1}{H}\frac{dX_7}{dt} &=& \frac{31   X_1^2 X_7}{60 X_5^2}-\frac{3}{10} X_2^2 X_7+\frac{3}{10}    X_3^2 X_7+\mu X_4+\frac{3}{10}    X_4^2 X_7\nonumber \\
    & & +\frac{1}{10}    X_5^2 X_7^3-\frac{13}{30}      X_6^2 X_7^3-\frac{7}{10} \lambda ^2    X_7 X_8^2\nonumber \\
    & & -\frac{7}{10} \lambda ^2    X_7^3+\frac{12 X_7}{5}, \label{eq:x7}
\\
    \frac{1}{H}\frac{dX_8}{dt} &=& \frac{31   X_1^2 X_8}{60 X_5^2}-\frac{3}{10} X_2^2 X_8+\mu X_3+\frac{3}{10}    X_3^2 X_8+\frac{3}{10}    X_4^2 X_8 \nonumber \\
    & & +\frac{1}{10}    X_5^2 X_7^2 X_8-\frac{13}{30}      X_6^2 X_7^2 X_8-\frac{7}{10} \lambda ^2    X_7^2 X_8\nonumber \\
    & & -\frac{7}{10} \lambda ^2    X_8^3+\frac{12 X_8}{5}, \label{eq:x8}
\\
    \frac{1}{H}\frac{d\mu}{dt} &=& \frac{12 \mu}{5}+\frac{31 \mu   X_1^2}{60 X_5^2}-\frac{3}{10} \mu X_2^2+\frac{3}{10}  \mu X_3^2+\frac{3}{10} \mu X_4^2 \nonumber \\
    & & +\frac{1}{10}  \mu X_5^2 X_7^2-\frac{13}{30}  \mu   X_6^2 X_7^2-\frac{7}{10} \lambda ^2   \mu X_7^2-\frac{7}{10} \lambda ^2   \mu X_8^2.
\end{eqnarray}
In order to find critical points from the dynamical system, we solve by setting all the derivatives, Eqs.(\ref{eq:x1}) - (\ref{eq:x8}) equal to zero together with the constraint equation (\ref{eq:const}). In addition, the critical point corresponds to the exact solution for each epoch of the dynamics of the universe. 

The stability of a critical point is analyzed by finding the eigenvalues of the matrix:
\begin{equation}
    M_{ij} = \frac{d}{dX_i}\left(\frac{1}{H}\frac{dX_j}{dt}\right).
    \label{stability-M}
\end{equation}
If all eigenvalues are negative, then a critical point is a stable fixed point. Whereas any positive eigenvalue signifies an instability of a critical point.

The energy densities for each type of field are following:
\begin{eqnarray}
    &&\Omega_{A} = -\frac{  X_1^2}{X_5^2} -     X_6^2 X_7^2,\qquad \Omega_{\phi} = X_2^2 +   X_5^2 X_7^2,
    \nonumber\\
    &&\Omega_{\eta_0} = - \left(X_3^2 + \lambda^2 X_8^2\right),\quad \Omega_{\eta_1} = - \left(X_4^2 + \lambda^2 X_7^2\right).
    \label{density}
\end{eqnarray}

The effective equation of state is given by
\begin{eqnarray}
w_{\rm eff} = -1 -\frac{2\dot{H}}{3 H^2}\,.
\label{EOS}
\end{eqnarray}
We can represent the equation of state in terms of the dimensionless variables by using in Eq.(\ref{Hdot}). 
We also can use the effective equation of state to identify the expansion phases where $w_{\rm eff} < -1/3$ and $w_{\rm eff} > -1/3$ represent the acceleration and deceleration expansion of the universe, respectively.
Moreover, the exact solution for each critical point can be obtained by performing the integration of the effective equation of state in Eq.(\ref{EOS}). One finds,
\begin{eqnarray}
a(t) = 
\begin{cases}
a_0 (t - t_0)^\gamma\,,\quad  w_{\rm eff} \neq  -1\,,\quad {\rm and} \qquad \gamma = \frac{2}{3(1+ w_{\rm eff})}\,,
\\
a_0 e^{C (t-t_0)}\,,\,\quad\; w_{\rm eff} = -1\,,
\end{cases}
\label{a-solution}
\end{eqnarray}
where $a_0$ and $t_0$ are integration constants of the scale factor and cosmic co-moving time respectively. For $w_{\rm eff} = 0$, the universe undergoes the matter-dominated phase, i.e., $a(t) \propto t^{2/3}$. For $w_{\rm eff} = -1$ with the positive constant, $C$, this is a dS critical point which corresponds to the acceleration expansion solution. It is worth noting that the exact solution of the scale factor is easily found for each critical point by substituting the effective equation of state. 

\section{Result}
Since the system involves 8 main parameters, $X_1 -X_8$, with 1 auxiliary parameter, $\mu$, it would be much easier to see the behaviours of the solution if the analysis is simplified somewhat. In this section we will first assume that there is no universal mass at 5 dimensional field theory $M_{(5)} = 0$. Then, we consider $M_{(5)} \neq 0$ in various corners of the phase space of the dynamical system.
\subsection{$M_{(5)} = 0$}
If $M_{(5)} = 0$, then $X_8$ is decoupled from the Friedman equation. It is easier to analyse with the assumption that $X_3 = 0$ and $X_8 = 0$ since $\frac{1}{H}\frac{dX_3}{dt} = 0$ and $\frac{1}{H}\frac{dX_8}{dt} = 0$ are satisfied automatically. After the analysis, we found that in the case, there is no real solution.
\subsection{$M_{(5)} \neq 0$}
It is evident that the complications of the autonomous system in Eqs.(\ref{eq:x1}-\ref{eq:x8}) in the case of $M_{(5)} \neq 0$ lead to the very complicated solutions of the fixed points which can not generically be represented in terms of $\lambda$ and $\mu$ parameters analytically. To demonstrate DM and DE profiles of the KK inspired BD model in this work, we simply analyze the dynamical system in our model by assuming that there are decoupled fields in various settings. However it is important to note that, according to the constraint equation in Eq.(\ref{eq:const}) and derivative of $X_2$ in Eq.(\ref{eq:x2}), we are forbidden to decouple $\phi$ (assuming $X_2 = 0$ along with $X_5 = 0$) for all cases.
\subsubsection{$(\phi, \eta_0)$-system}
We first consider the $(\phi, \eta_0)$-system case i.e., $X_3 \neq 0$ and $X_8 \neq 0$ due to the density parameters defined in Eq.(\ref{density}) and the constraint equation in Eq.(\ref{eq:const}).  
In this case, we obtain,
\begin{eqnarray}
 {X}_1 = X_4 =  {X}_6 = X_7 = 0 \,.
\end{eqnarray}
Moreover, the constraint equation can be rewritten as
\begin{eqnarray}
1 = X_2^2 - X_3^2 - \lambda^2 X_8^2\,.
\end{eqnarray}
This constraint reduces the autonomous system down to 3 dimensional phase space as
\begin{eqnarray}
    \frac{1}{H}\frac{dX_3}{dt} &=& -\frac{3}{10} \left( 1 + X_3^2 + \lambda^2 X_8^2 \right) X_3-\sqrt{\frac{3}{2}} X_3\sqrt{1 +  X_3^2 +  \lambda^2 X_8^2 } 
    \nonumber \\
    & & -\frac{7}{10} \lambda ^2 X_3 X_8^2
    +\frac{3}{10} X_3^3-\frac{3 X_3}{5}-\lambda ^2 \mu X_8,
    \\
    \frac{1}{H}\frac{dX_5}{dt} &=& -\sqrt{\frac{3}{2}} X_5 \sqrt{1 + X_3^2 +  \lambda^2 X_8^2 },
    \\
    \frac{1}{H}\frac{dX_8}{dt} &=& \mu X_3+\frac{3}{10}  X_3^2 X_8 -\frac{7}{10} \lambda ^2  X_8^3+\frac{12 }{5}X_8 \,.
\end{eqnarray}
In this case, we found that all real positive fixed points of this autonomous system always come with the $X_5 =0$ solution. This means the dilaton field, $\phi$ diverges $\phi\to\infty$ which is an unphysical solution. Therefore we do not further analyze the case where $\eta_0$ is not decoupled, i.e., we will only assume $X_3 = X_8 = 0$ from now on. 


\subsubsection{$(\phi, \eta_1)$-system}
In this case, we consider $(\phi, \eta_1)$-system, $X_4\neq 0$ and $X_7\neq 0$. This leads to,
\begin{eqnarray}
  X_1 = X_3 =   X_6 = X_8 = 0\,,
\end{eqnarray}
with the following constraint equation,
\begin{eqnarray}
1 = X_2^2  - X_4^2 +  X_5^2 X_7^2 - \lambda^2 X_7^2\,.     
\end{eqnarray}
The autonomous system is reduced to three differential equations as
\begin{eqnarray}
    \frac{1}{H}\frac{dX_4}{dt} &=& -\frac{3}{10}\left( 1 + X_4^2 - X_5^2 X_7^2 +  \lambda^2 X_7^2\right) X_4 +\frac{3}{10}   X_3^2 X_4 \nonumber \\
    & & +\frac{1}{10}  X_4 X_5^2 X_7^2-\sqrt{\frac{3}{2}} X_4\sqrt{1 +  X_4^2 -  X_5^2 X_7^2 +  \lambda^2 X_7^2}  \nonumber \\
    & & -\frac{7}{10} \lambda ^2   X_4 X_7^2 +\frac{3}{10}   X_4^3-\frac{3 X_4}{5} +\mu X_5^2 X_7 -\lambda ^2 \mu X_7, 
\\
    \frac{1}{H}\frac{dX_5}{dt} &=& -\sqrt{\frac{3}{2}} X_5 \sqrt{1 +  X_4^2 -  X_5^2 X_7^2 +  \lambda^2 X_7^2},
\\
    \frac{1}{H}\frac{dX_7}{dt} &=& -\frac{3}{10} \left( 1 + X_4^2 - X_5^2 X_7^2 + \lambda^2 X_7^2\right) X_7 +\mu X_4+\frac{3}{10}   X_4^2 X_7\nonumber \\
    & & +\frac{1}{10}   X_5^2 X_7^3 -\frac{7}{10} \lambda ^2   X_7^3+\frac{12 X_7}{5}\,.
\end{eqnarray}
The critical points in this case are given by
\begin{eqnarray}
X_4 &=& \frac{\Delta}{408 \mu} \left(- 75 + \sqrt{816 \lambda ^2 \mu^2+5625}\right),
\\
X_5 &=& \frac{1}{34 \mu^2}\sqrt{1360 \lambda ^2 \mu ^4-\frac{27}{2} \mu ^2 \left(- 75 +\sqrt{ 816 \lambda ^2 \mu ^2+5625}\right)},
\\
X_7 &=& \frac{\Delta}{6},
\\
\Delta^2 &=& \frac{\sqrt{816 \lambda ^2 \mu ^2+5625} + 75}{\lambda ^2}\,.
\end{eqnarray}
\begin{figure}
    \centering
    \includegraphics[width=10cm]{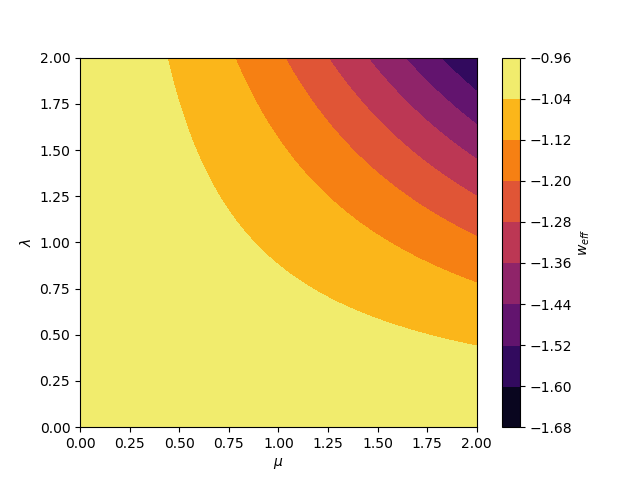}
    \caption{The contour plot for the equation of state in the case where $\eta_1$ is not decoupled.}
    \label{fig:w2epp}
\end{figure}
The effective equations of state in the $\eta_1 \neq 0$ case are given by
\begin{eqnarray}
w_{\rm eff}&=& \frac{1}{102} \left(-27 - \sqrt{816 \lambda ^2 \mu ^2+5625}\right).
\end{eqnarray}
The equation of state as a function of $\mu$ is depicted in Fig. \ref{fig:w2epp}. As a result, we found that $w_{\rm eff} \leq -1$.  In addition, the non-vanishing energy density parameters in this case are given by 
\begin{eqnarray}
\Omega_{\phi} &=& \frac{5}{306} \left(2 \sqrt{816 \lambda ^2 \mu^2 +5625} + 105\right), \\
\Omega_{\eta_1} &=& \frac{1}{306} \left(- 831-10 \sqrt{816 \lambda ^2 \mu^2+5625}\right).
\end{eqnarray}
Having use the definition of the stability matrix in Eq.(\ref{stability-M}), the real parts of the eigenvalues in this case are always coming with the plus and minus signs for positive $\lambda$ and $\mu$. This shows that these critical points are saddle points. Therefore, $\eta_1$ and $\phi$ are insufficient to exhibit DM/DE behaviour. 

\subsubsection{$(\phi, \eta_1,  {A})$-system}
\begin{figure}
    \centering
    \includegraphics[width=10.5cm]{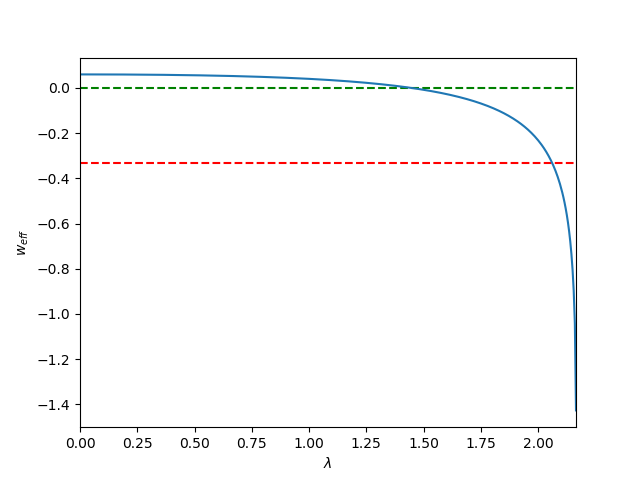}
    \caption{The equation of state for $(\phi, \eta_1,  {A})$-system. Notice that depending on the value of $\lambda$, the system is able to exhibit DM and DE bahaviour. The green and red dashed lines are $w_{\rm eff} = 0$ and $w_{\rm eff}=-1/3$, respectively.  }
    \label{fig:w4}
\end{figure}

The final consideration in this section is non-vanishing $ {\mathcal{A}}_c$ field. According to the density parameter of the $ {\mathcal{A}}_c$ field, we impose,
\begin{eqnarray}
X_3 = X_4 = X_8 =0\,.
\end{eqnarray}
Notice that we further simplify the system by assume that the kinetic term of $\eta_1$ ($X_4$) vanishes. The constraint equation in this case is written by
\begin{eqnarray}
1 = -\frac{ {X}_1^2}{2X_5^2} + X_2^2 + \left(X_5^2 -  {X}_6^2 \right)X_7^2 - \lambda^2 X_7^2 \,. 
\end{eqnarray}
The autonomous system for this case is composed of four first order differential equations as
\begin{eqnarray}
\frac{1}{H}\frac{d   X_1}{dt} &=& -\frac{3}{10}   X_1 X_2^2 + \frac{1}{10}   X_1 X_5^2 X_7^2 + \frac{2   X_1}{5} -    X_6 X_7^2 -\frac{13}{30}    X_1   X_6^2 X_7^2\nonumber \\
    & & -\frac{7}{10}  \lambda^2   X_1 X_7^2  + \frac{31   X_1^3}{60 X_5^2},
\\
    \frac{1}{H}\frac{dX_2}{dt} &=& \frac{31   X_1^2 X_2}{60 X_5^2}+\frac{2 \sqrt{6}   X_1^2}{5 X_5^2} +\frac{1}{10}  X_2 X_5^2 X_7^2 -\frac{13}{30}  X_2   X_6^2 X_7^2\nonumber \\
    & & -\frac{7}{10} \lambda ^2  X_2 X_7^2 -\frac{1}{10} 3 X_2^3-\frac{7}{5} \sqrt{\frac{3}{2}} X_2^2 -\frac{3 X_2}{5}\nonumber \\
    & & +\frac{2}{5} \sqrt{6}   X_5^2 X_7^2-\frac{2}{5} \sqrt{6}     X_6^2 X_7^2 -\frac{4}{5} \sqrt{6} \lambda ^2   X_7^2 +\frac{3 \sqrt{6}}{5}, 
\\
    \frac{1}{H}\frac{dX_5}{dt} &=& -\sqrt{\frac{3}{2}} X_2 X_5,
\\
    \frac{1}{H}\frac{d  X_6}{dt} &=&   X_1  -   X_6,
\end{eqnarray}
where the $X_7$ variable will be replaced by
\begin{eqnarray}
X_7^2 = \frac{  X_1^2 +2 \left(1 - X_2^2\right) X_5^2}{2 X_5^2 \left(\lambda ^2 - X_5^2 +   X_6^2\right)}\,.
\end{eqnarray}
The critical points in this case are read
\begin{eqnarray}
\mathcal{C}_1\,:~ &&   X_1 = 0,~ X_2 = 0,~ X_5 = \sqrt{\frac{7}{5}} \lambda,~   X_6 = 0
\\
\mathcal{C}_2\,:~ &&   X_1 = \frac{\sqrt{ 4 \sqrt{-39 \lambda ^4+60 \lambda ^2+576} -54 \lambda ^2 -81}}{\sqrt{59}},~ X_2 = 0,
\nonumber\\
&& X_5 = \frac{\sqrt{53 \lambda ^2+7 \sqrt{-39 \lambda ^4+60 \lambda ^2+576}-186}}{\sqrt{118}},~ 
\nonumber\\
&&   X_6 = \frac{\sqrt{ 4 \sqrt{-39 \lambda ^4+60 \lambda ^2+576} -54 \lambda ^2 -81}}{\sqrt{59}}\,.
\end{eqnarray}
The equation of state is given by
\begin{eqnarray}
w_{\rm eff} &=& \frac{1}{45 X_5^2 \left(X_5^2-  X_6^2-\lambda ^2\right)}\Big[   X_1^2 \left(-26 \lambda ^2+17 X_5^2-22   X_6^2\right)
\nonumber\\
&+& 2 X_5^2 \left(X_2^2 \left(15 \lambda ^2-6 X_5^2+11   X_6^2\right)+15 X_5^2-20   X_6^2-24 \lambda ^2\right)\Big].
\label{eos-3}
\end{eqnarray}
For the $\mathcal{C}_1$ critical point, we obtain,
\begin{equation}
w_{\rm eff}^{(1)} = -\frac13\,
\end{equation}
which sits on the border between accelerating and decelerating universe.
The non-vanishing energy density parameters for $\mathcal{C}_1$ are read
\begin{eqnarray}
\Omega_\phi^{(1)} = \frac72\,,\qquad \Omega_{\eta_1}^{(1)}= -\frac52\,.
\end{eqnarray}
While for the $\mathcal{C}_2$ critical point, the $w_{\rm eff}$ takes the complicated form by substituting $\mathcal{C}_2$ in Eq.(\ref{eos-3}). It reads,
\begin{eqnarray}
w_{\rm eff}^{(2)} &=& \frac{(9926-248 \alpha ) \lambda ^2-2614 \alpha -2552 \lambda ^4+62580}{3 \left(\alpha -43 \lambda ^2+24\right) \left(7 \alpha +53 \lambda ^2-186\right)},
\\
\alpha &=& \sqrt{-39 \lambda ^4+60 \lambda ^2+576}\,.\nonumber
\end{eqnarray}
To illustrate this, we therefore plot the $w_{\rm eff}$ as a function of $\lambda$ and it is shown in Fig. \ref{fig:w4}. At $\lambda = 1.45$, this gives $w_{\rm eff} =0$ representing DM phase. We note that at $\lambda = 2.062$ is the crossing point of the equation of state for DE. In addition, we found the non-vanishing energy density parameters for $\mathcal{C}_2$ as
\begin{eqnarray}
\Omega_{  A_c}^{(2)} &=& \frac{4 \left(4 \xi -54 \lambda ^2 -81\right) \left(7 \xi -6 \lambda ^2 -186\right)}{\left(\xi -43 \lambda ^2 + 24\right) \left(7\xi + 53 \lambda ^2 -186\right)},
\\
\Omega_\phi^{(2)} &=& \frac{\left(\xi + \lambda ^2 -12\right) \left( 7 \xi+ 53 \lambda ^2 -186\right)}{2 \left(9- 2 \lambda ^2\right) \left(\xi -43 \lambda ^2 +24\right)},
\\
\Omega_{\eta_1}^{(2)} &=& \frac{59 \lambda ^2 \left(\lambda ^2+\xi-12\right)}{\left(2 \lambda ^2-9\right) \left(\xi-43 \lambda ^2 +24\right)},
\\
\xi &=& \sqrt{576 + 60 \lambda ^2 - 39 \lambda ^2}\,.
\end{eqnarray}
The eigenvalues (real part) of the stability matrix for $\mathcal{C}_1$ are read
\begin{equation}
\big(-1,~-1,~-1,~-1\big)\,.
\end{equation}
We find $\mathcal{C}_1$ being stable points.
On the one hand, the eigenvalues (real part) of the stability matrix for $\mathcal{C}_2$ with $\lambda = 1.45$ (DM phase i.e., $w_{\rm eff} = 0$) are given by
\begin{eqnarray}
\big(7.395,~ -1.995,~ -1.995,~ 1.204\big)\,,
\end{eqnarray}
which means saddle point. On the other hand, the eigenvalues (real part) for $\mathcal{C}_2$ with $\lambda = 2.16$ (DE phase with the dS solution, $w_{\rm eff} = -1$) are found as
\begin{eqnarray}
\big(-0.472,~ -0.472,~ -3.292,~ -0.204\big)\,,
\end{eqnarray}
which means stable (attractor) point. Therefore, in this $(\phi, \eta_1,  {A})$-system, DM phase and DE phase are present depending on the $\lambda$ values.

\section{Conclusion and Outlook}
In this work, we have presented the KK inspired BD model in order to solve DM \& DE problem. We start from the traditional KK gravitational action with an introduction of the scalar field and 2 gauge fields in the 5 dimensional spacetime. The KK compactification process makes the dilaton coupling to the scalar fields in the bulk spacetime giving a particular form of the potential. The gauge field from 5 dimensional metric and 2 additional gauge fields can form the mutually orthogonal spatial vectors leading to consistency with the isotropic universe. We continue to derive the equations of motion for all relevant matters. The dynamical system is utilized to demonstrate the existence of DM \& DE in the model. The autonomous system of equations, the effective equation of the state and the exact solution of the scale factor are constructed. Due to its intricacy of the dynamical system, the qualitative analysis has been done separately for $M_{(5)} = 0$ and $M_{(5)} \neq 0$ cases in order to aid our understanding.

For the $M_{(5)}= 0$, we have found that there is no real solution of the critical points in this case. We conclude that there is no solution corresponding to DM \& DE. Rich phenomena of cosmological states can be achieved when the mass parameter of the scalar field, $M_{(5)}$ is turned on. We extensively divide the critical points analysis into three sub cases as $(\phi, \eta_0)$, $(\phi, \eta_1)$, $(\phi, \eta_1, \mathcal{{A}}_c)$ systems. It is worth mentioning that we do not decouple the dilaton from the system to prevent the divergence of the dilaton which could lead to an unphysical solution. The interesting physics of the critical points is in the $(\phi, \eta_1, \mathcal{{A}}_c)$ system. There is a critical point that exhibits both DM ($w_{\rm eff} =0$) \& DE ($w_{\rm eff}=-1$) behavior depending on the $\lambda$ parameter. At the DM phase, on the one hand, the critical point has the scaling solution, $a \propto t^{2/3}$ and it is a saddle point. In DE phase, on the other hand, we found a stable critical point with the dS solution, $a \propto e^{C t}$ when the $\lambda$ parameter increasing to a particular value. The results suggest that $\lambda = M_{(5)}R_k$ is crucial to the KK inspired BD model in the sense that it controls the DM \& DE behavior. Nevertheless, the presence of the gauge field is also important to the system as the only case with DM \& DE existences has the gauge field couple to the system. 

The unified models of DM/DE, according to the literature survey, are interesting and active topics in cosmology. The Chaplygin gas model is the most popular as a description of DM/DE in a single framework \cite{Kamenshchik:2001cp,Bilic:2001cg,Bento:2002ps,Sandvik:2002jz,Bento:2004uh}. In addition, various models of scalar fields and k-essence are also used as the unified DM/DE model. \cite{Padmanabhan:2002sh,Scherrer:2004au,Bilic:2008yr,Bertacca:2007ux,Ansoldi:2012pi,Guendelman:2015jii,Koutsoumbas:2017fxp,Brandenberger:2018xnf,Brandenberger:2019jfh}. Moreover, BSM physics and string theory also provide alternative solution for DM/DE models as well \cite{Mainini:2005mq,Hung:2005ft,Arefeva:2005ixb,Takahashi:2005kp,Liddle:2006qz,Bertolami:2007wb,Brandenberger:2020gaz}. In contrast to the inclusion of the exotic matter as DM/DE models, the modified theories of gravity and the geometrical effects are widely used as the explanation of the unified DM/DE models \cite{Kim:2004is,Mannheim:2005bfa,Capozziello:2005tf,Capozziello:2006uv,Banados:2008rm,Dehghani:2008xf,Benisty:2018qed,Anagnostopoulos:2019myt}. Although, all those unified DM/DE models mentioned above can be made to be compatible with current observational data. However, the model present in the work is a new alternative approach of the unified DM/DE framework. The merit of this approach is the strong connection to new physics coming from extra dimension. Particularly, our model can provide additional potential detection channel of new physics via the extra gauge field sector. For example, one could study the connection with the dark photon constraints. We consider this work as a promising toy model. To gain a deeper understanding of the problems, the study in a more realistic set up is required for future work and further investigation, e.g., inclusion of the barotropic fluid, extraction of the observables from the model and comparing them to the cosmological data.

\section*{Acknowledgements} \label{sec:acknowledgements}
The work of CP has been supported in part by the Thailand Research Fund under contract No.MRG6280131 and by the  National Astronomical Research Institute of  Thailand. DS is supported by Thailand Research Fund (TRF) under a contract No.TRG6180014.


\bibliographystyle{elsarticle-num}

\bibliography{KK-BD}


\end{document}